\documentclass[doublecol]{epl2} 
\usepackage{graphicx}
\usepackage{amsmath}
\usepackage{amssymb}
\usepackage{bm}
\title{Anomalously large conductance fluctuations in weakly disordered graphene}
\author{A. Rycerz\inst{1} \and J. Tworzyd{\l}o\inst{2} \and C.W.J. Beenakker\inst{3}}
\shortauthor{A. Rycerz \etal}
\institute{
\inst1{Marian Smoluchowski Institute of Physics, Jagiellonian University, Reymonta 4, 30--059 Krak\'{o}w, Poland}\\
\inst2{Institute of Theoretical Physics, Warsaw University, Ho\.{z}a 69, 00--681 Warsaw, Poland}\\
\inst3{Instituut-Lorentz, Universiteit Leiden, P.O. Box 9506, 2300 RA Leiden, The Netherlands}
}
\pacs{73.23.-b}{Electronic transport in mesoscopic systems}
\pacs{73.20.Fz}{Weak or Anderson localization}
\pacs{73.40.-c}{Electronic transport in interface structures}
\abstract{
We have studied numerically the mesoscopic fluctuations of the conductance of a graphene strip (width $W$ larger than length $L$), in an ensemble of samples with different realizations of the random electrostatic potential landscape. For strong disorder (potential fluctuations comparable to the hopping energy), the variance of the conductance approximates the value predicted by the Altshuler-Lee-Stone theory of universal conductance fluctuations, ${\rm Var}\,G_{\rm UCF}=0.12\,(W/L)(2e^2/h)^2$. For weaker disorder the variance is greatly enhanced if the potential is smooth on the scale of the atomic separation. There is no enhancement if the potential varies on the atomic scale, indicating that the absence of backscattering on the honeycomb lattice is at the origin of the anomalously large fluctuations.
}

\begin{document}
\maketitle

\begin{figure}[tb]
\centerline{\includegraphics[width=0.9\linewidth]{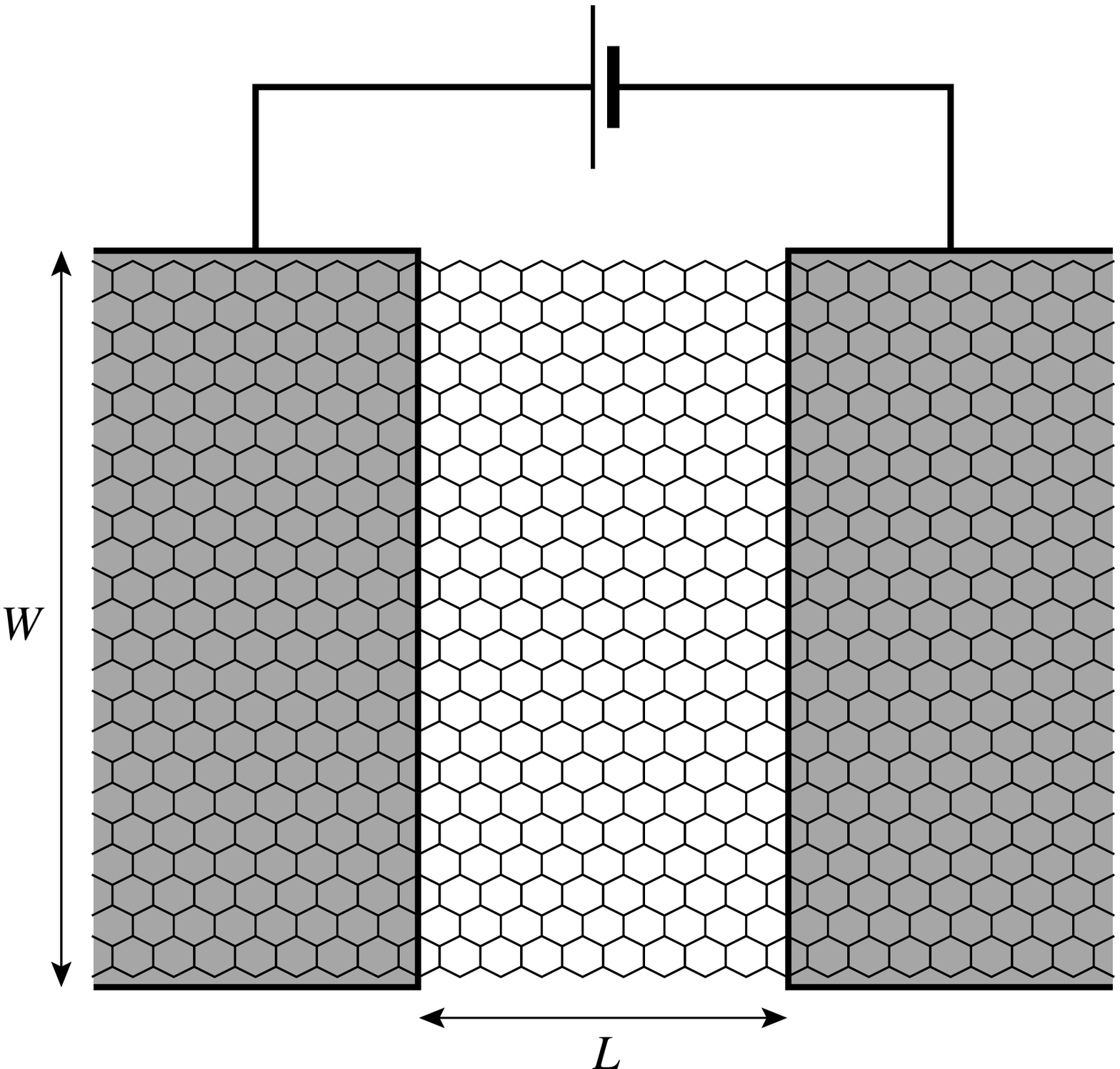}}

\centerline{\includegraphics[width=0.9\linewidth]{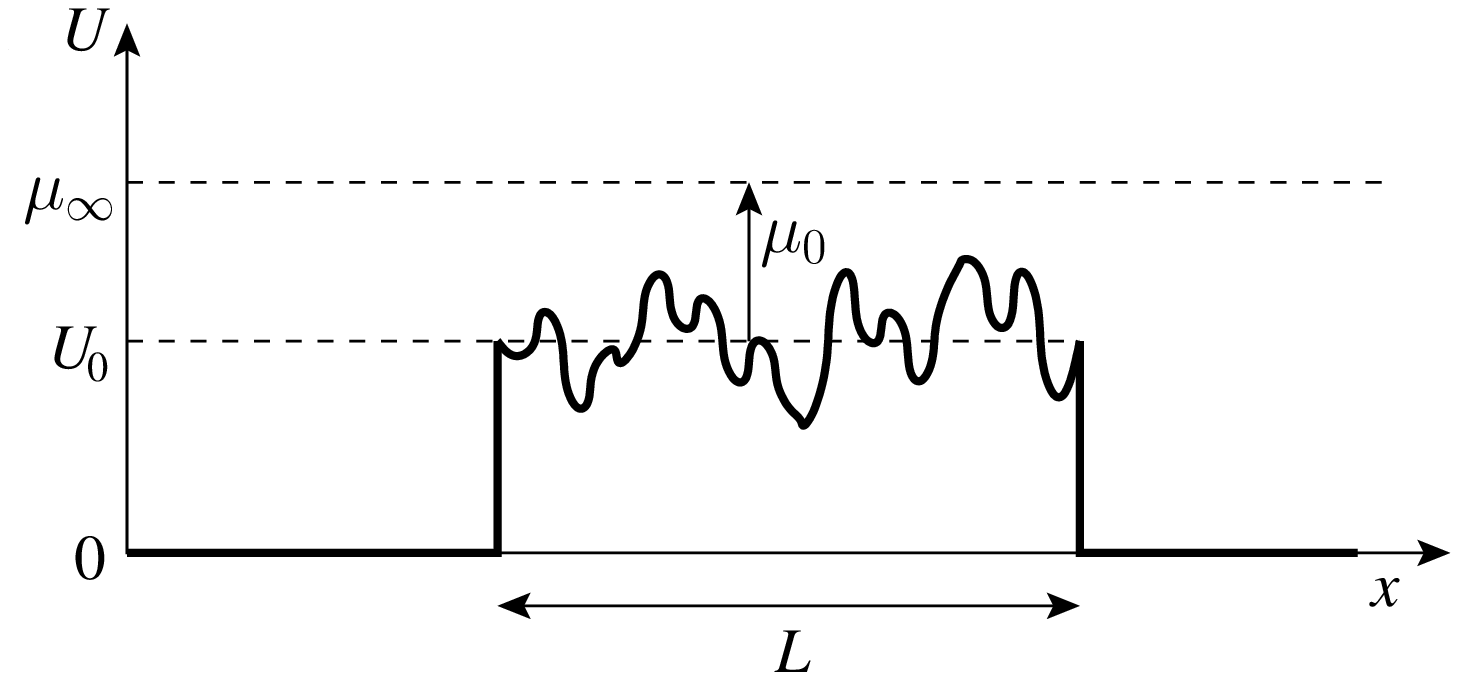}}
\caption{\label{fig_device}
Top panel: Top view of the honeycomb lattice in a graphene strip, connecting two electrical contacts at a voltage difference (gray rectangles). The samples used in the simulation are much larger than the one shown here. Bottom panel: Potential profile along the strip, showing the fluctuations from the disorder.
}
\end{figure}

Phase coherent diffusion in metals is accompanied by sample-to-sample fluctuations in the conductance of the order of the conductance quantum $e^{2}/h$, dependent on the shape of the conductor but independent of its size or of the disorder strength. This is the phenomenon of universal conductance fluctuations (UCF) \cite{Alt85,Lee85}. The universality does not extend to different transport regimes, in particular the fluctuations become much {\em smaller\/} than the UCF value both in the ballistic regime of weak disorder and in the localized regime of strong disorder \cite{Bee91,Akk06}.

In this paper we report on the observation in a computer simulation of a transport regime with conductance fluctuations that are much {\em larger\/} than the UCF value. The anomalously large fluctuations appear in a tight-binding model of a carbon monolayer, for a disorder potential that is smooth on the scale of the atomic separation and weak on the scale of the hopping energy. It is known that such a potential in a honeycomb lattice can deflect the electrons but cannot scatter them backwards \cite{And98,Sho98}. The consequences for weak localization of the absence of backscattering have been studied theoretically \cite{Suz02,Khv06,McC06,Mor06a} and experimentally \cite{Mor06b,Wu06}. While conductance fluctuations as a function of magnetic field in a given sample have been observed experimentally \cite{Mor06b,Ber06,Hee06}, and analyzed in terms of the UCF theory, the anomaly found here in the sample-to-sample fluctuations has not been reported previously.

We consider a disordered strip of graphene in the $x-y$ plane, connected to ballistic leads at $x=0$ and $x=L$ (see Fig.\ \ref{fig_device}). The orientation of the honeycomb lattice is such that the edges at $y=0$ and $y=W$ are in the zigzag configuration. We vary $L$ and $W$ at fixed aspect ratio (mostly taking a rather large ratio $W/L=3$ to minimize the effects of edge states). The lattice Hamiltonian is
\begin{equation}\label{Hdef}
H=\sum_{i,j}\tau_{ij}|i\rangle\langle j|+\sum_{i}\bigl[U_{\rm gate}(\bm{r}_{i})+U_{\rm imp}(\bm{r}_{i})\bigr]|i\rangle\langle i|.
\end{equation}
The hopping matrix element $\tau_{ij}=-\tau$ if the orbitals $|i\rangle$ and $|j\rangle$ are nearest neigbors (with $\tau\approx3\,{\rm eV}$), otherwise $\tau_{ij}=0$. The velocity $v$ near the Dirac point equals $v=\tfrac{1}{2}\sqrt{3}\,\tau a/\hbar\approx 10^{6}\,{\rm m/s}$, with $a=0.246\,{\rm nm}$ the lattice constant.

The electrostatic potential contains a contribution $U_{\rm gate}$ from gate electrodes and a random contribution $U_{\rm imp}$ from impurities. The potential $U_{\rm gate}$ vanishes in the leads $x<0$ and $x>L$ and equals $U_{0}$ in the strip $0<x<L$. By varying $U_{0}$ at fixed Fermi energy $\mu_{\infty}$ in the leads, we can vary the Fermi energy $\mu_{0}=\mu_{\infty}-U_{0}$ in the strip. We take $\mu_{\infty}=\tau/2$ and compare the two cases $U_{0}=\mu_{\infty}\Rightarrow\mu_{0}=0$ and $U_{0}=0\Rightarrow\mu_{0}=\mu_{\infty}$. The first case is an undoped graphene strip, the second case is a heavily doped strip (but still at sufficiently small Fermi energy that the linearity of the dispersion relation holds reasonably well).

We generate a realization of the disorder potential by randomly choosing $N_{\rm imp}$ lattice sites $\bm{R}_{1},\bm{R}_{2},\ldots \bm{R}_{N_{\rm imp}}$ out of the total number $N_{\rm tot}=\frac{4}{3}\sqrt{3}\,LW/a^{2}$ of sites in the disordered strip, and by randomly choosing the potential amplitude $U_{n}$ at the $n$-th site in the interval $(-\delta,\delta)$. We then smooth the potential over a range $\xi$ by convolution with a Gaussian,
\begin{equation}
U_{\rm imp}(\bm{r})=\sum_{n=1}^{N_{\rm imp}}U_{n}\exp\left(-\frac{|\bm{r}-\bm{R}_{n}|^{2}}{2\xi^{2}}\right).\label{Uimpdef}
\end{equation}
In the special case $\xi\ll a$, $N_{\rm imp}=N_{\rm tot}$ each of the lattice sites in the strip has a randomly fluctuating potential. This is the Anderson model on a honeycomb lattice studied in Ref.\ \cite{Ver06}. We contrast this model of atomic-scale defects with the case $\xi=a\sqrt{3}$ of a potential which is still short-ranged on the scale of the system size but which varies smoothly on the atomic scale. Such a potential could be realized by screened charges in the substrate. (The Gaussian smoothing is chosen for computational convenience, and we have checked that the results are not sensitive to the type of smoothing considered.) 

We quantify the disorder strength by the dimensionless correlator
\begin{equation}
K_{0}=\frac{LW}{(\hbar v)^{2}}\frac{1}{N_{\rm tot}^{2}}\sum_{i=1}^{N_{\rm tot}}\sum_{j=1}^{N_{\rm tot}}\langle U_{\rm imp}(\bm{r}_{i})U_{\rm imp}(\bm{r}_{j})\rangle\label{K0def}
\end{equation}
of the random impurity potential (with vanishing average, $\langle U_{\rm imp}\rangle=0$). This single number $K_{0}$ is representative on length scales large compared to the correlation length $\xi$. For the model potential (\ref{Uimpdef}) we find (for $\xi\ll L,W$)
\begin{eqnarray}
K_{0}&=&\tfrac{1}{9}\sqrt{3}\,(\delta/\tau)^{2}(N_{\rm imp}/N_{\rm tot})\kappa^{2},\label{K0kappa}\\
\kappa&=&\frac{1}{N_{\rm imp}}\sum_{n=1}^{N_{\rm imp}}\sum_{i=1}^{N_{\rm tot}}\exp\left(-\frac{|\bm{r_{i}}-\bm{R}_{n}|^{2}}{2\xi^{2}}\right)\nonumber\\
&=&\left\{\begin{array}{cl}
1&{\rm if}\;\;\xi\ll a,\\
\frac{8}{3}\pi\sqrt{3}\,(\xi/a)^{2}&{\rm if}\;\;\xi\gg a.
\end{array}\right.\label{kappadef}
\end{eqnarray}

For large $\mu_{0}$ the correlator $K_{0}$ determines the transport mean free path $l_{\rm tr}$ in Born approximation \cite{Sho98,Suz02},
\begin{equation}
l_{\rm tr}=\frac{2}{k_{F}K_{0}}\times\left\{
\begin{array}{ll}
2&{\rm if}\;\;\xi\gtrsim a,\\
1&{\rm if}\;\;\xi\ll a,
\end{array}\right.\label{ltrdef}
\end{equation}
where $k_{F}=|\mu_{0}|/\hbar v$ is the Fermi wave vector in the strip (which should be $\gg 1/l_{\rm tr}$ for the Born approximation to hold). The factor-of-two increase in $l_{\rm tr}$ for smooth disorder is due to the absence of backscattering in the honeycomb lattice \cite{And98,Sho98}. The corresponding ``classical'' conductivity (without quantum corrections) is given by $\sigma_{\rm class}=(2e^{2}/h)k_{F}l_{\rm tr}$.

\begin{figure}[tb]
\centerline{\includegraphics[width=0.9\linewidth]{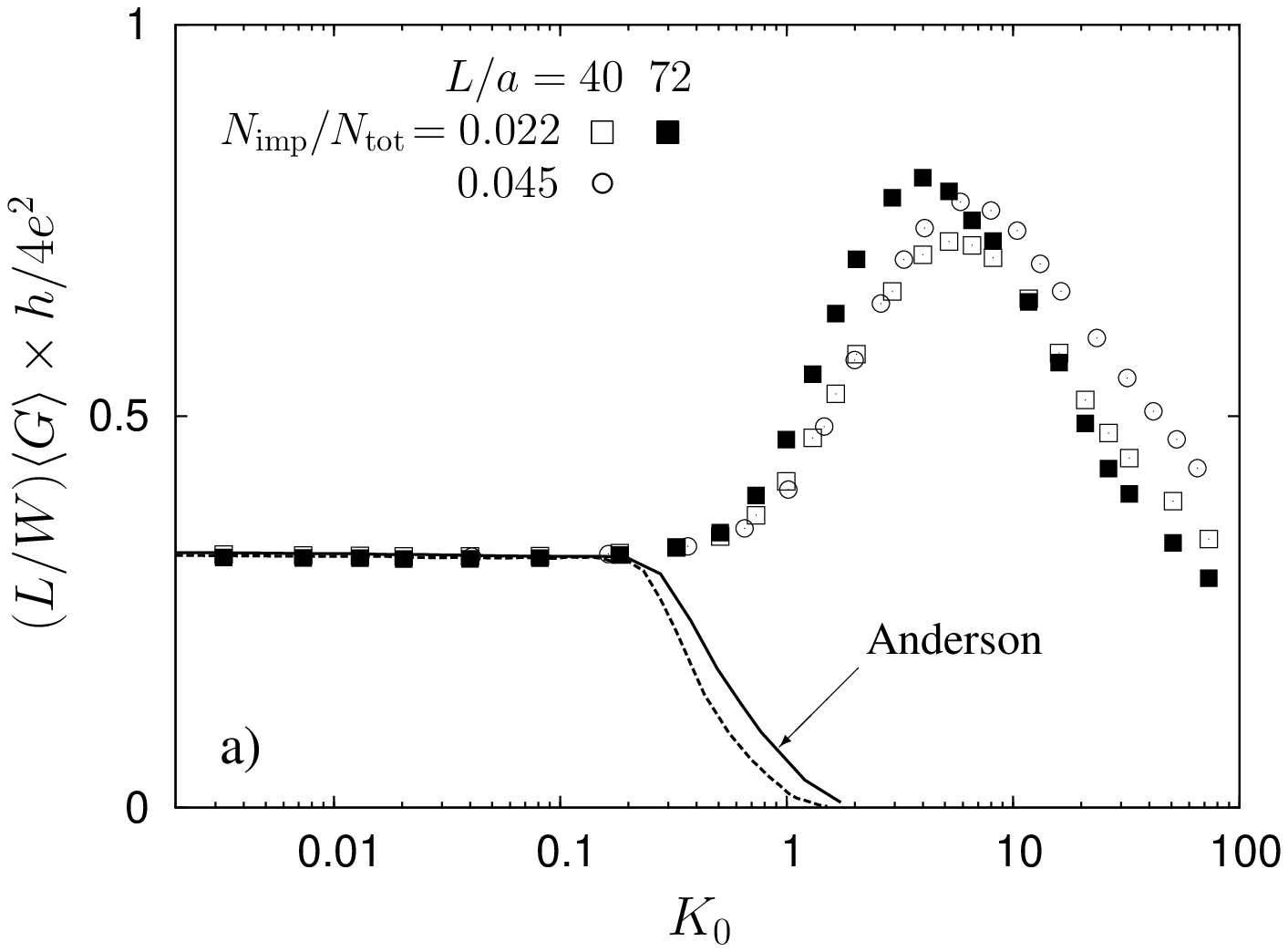}}

\centerline{\includegraphics[width=0.9\linewidth]{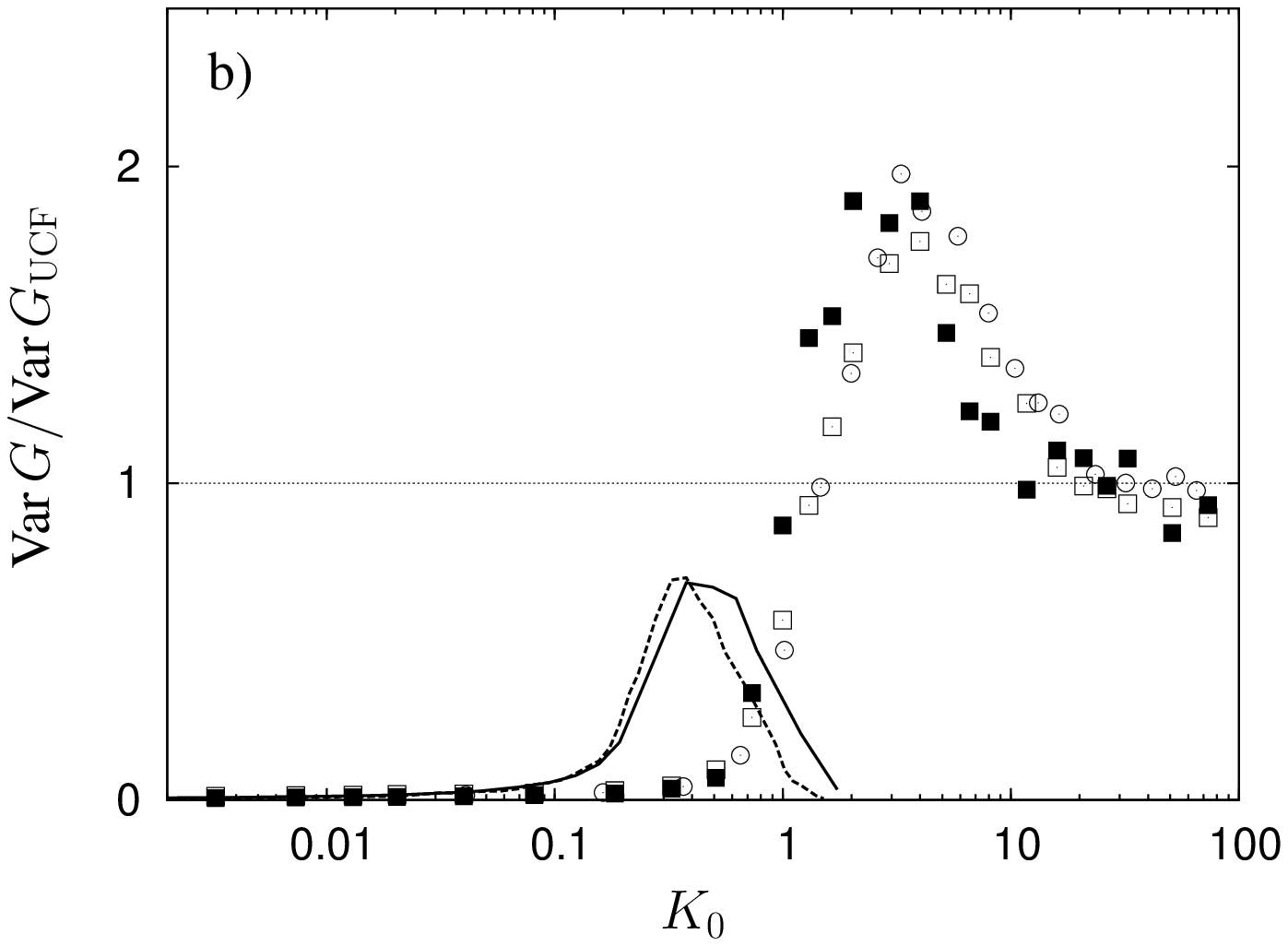}}
\caption{\label{fig_mu00}
Average and variance of the conductance as a function of the strength of the disorder potential, quantified by the correlator (\ref{K0def}). These plots are for the case that the disordered strip is at the Dirac point ($\mu_{0}=0$). The data points are for a smooth, short-range impurity potential (correlation length $\xi=a\sqrt{3}$), with different values of the impurity density $N_{\rm imp}/N_{\rm tot}$. Open symbols are for $L=40\,a$, filled symbols for $L=72\,a$ (at fixed aspect ratio $W/L=3$). The solid and dashed lines are the Anderson model of atomic scale disorder [$\xi=0$, $N_{\rm imp}=N_{\rm tot}$, $L=40\,a$ (solid) and $L=72\,a$ (dashed)].
}
\end{figure}

\begin{figure}[tb]
\centerline{\includegraphics[width=0.9\linewidth]{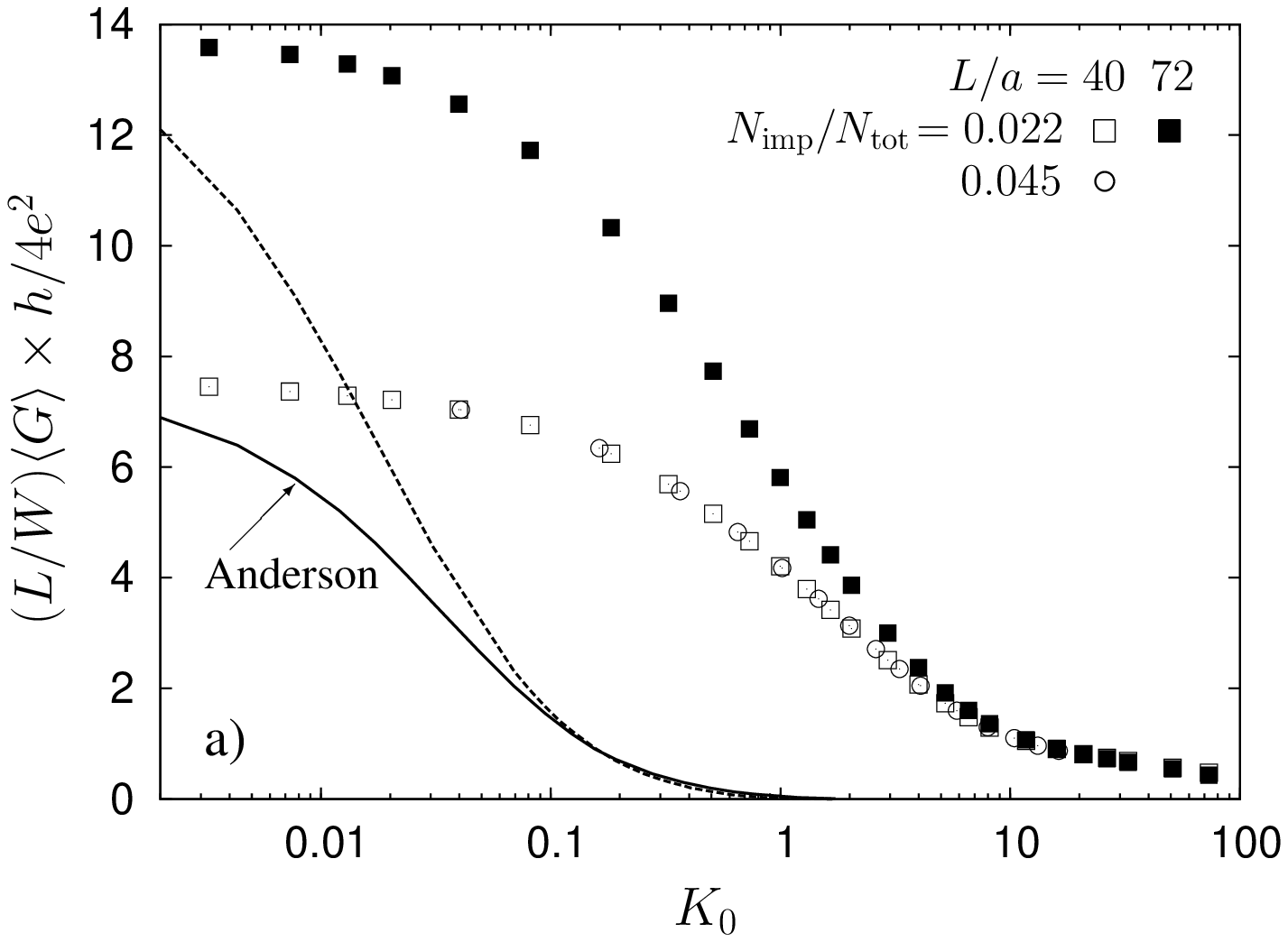}}

\centerline{\includegraphics[width=0.9\linewidth]{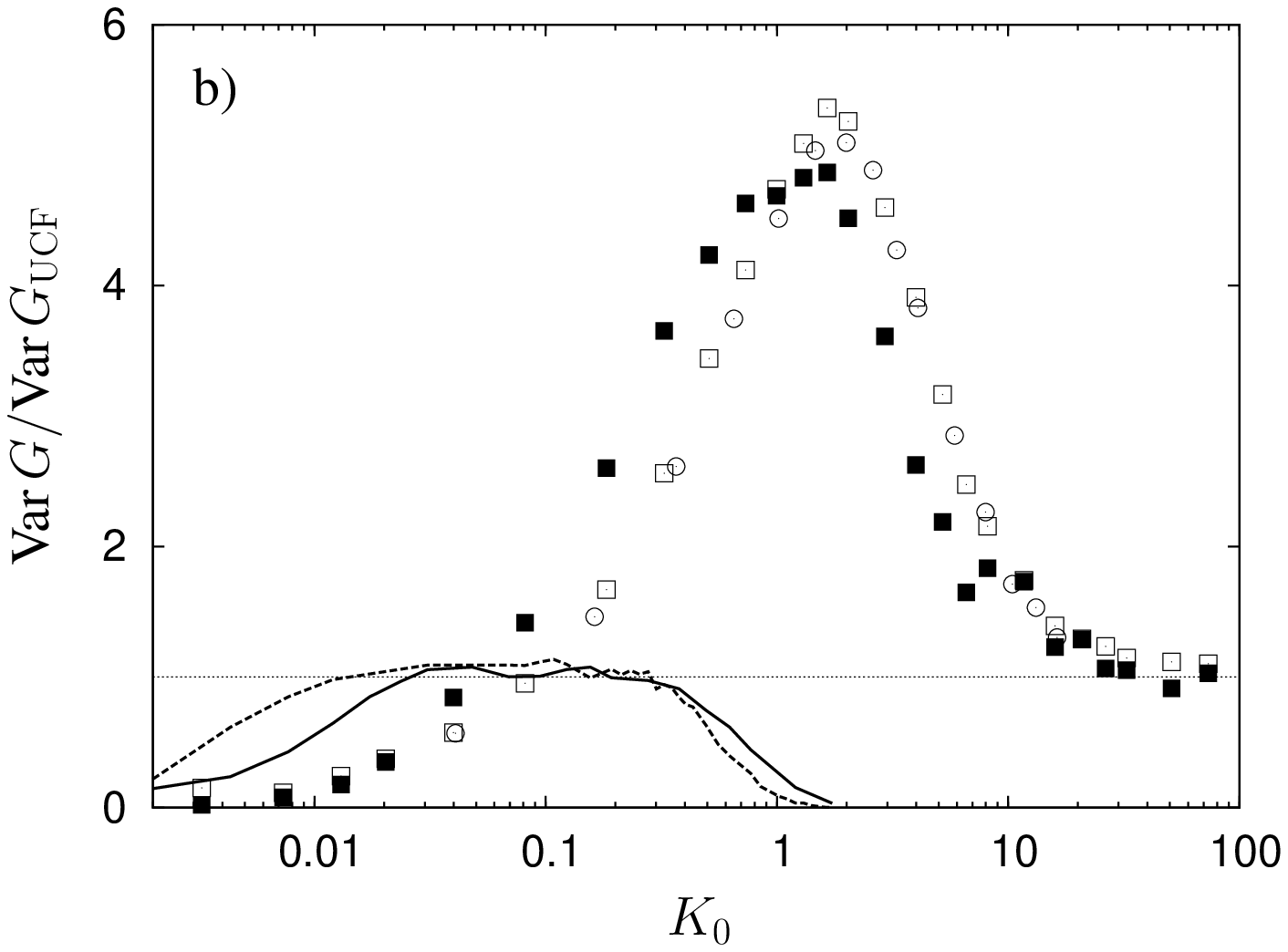}}
\caption{\label{fig_mu05}
Same as Fig.\ \ref{fig_mu00}, but now for the case that the disordered strip is away from the Dirac point ($\mu_{0}=\tau/2=\mu_{\infty}$).
}
\end{figure}

We calculate the transmission matrix $\bm{t}$ numerically by means of a recursive Green function algorithm. The conductance $G$ then follows from the Landauer formula $G=(2e^{2}/h){\rm Tr}\,\bm{t}\bm{t}^{\dagger}$. (The factor of two accounts for the spin degeneracy.) By repeating the calculation for some 300--3000 realizations of the disorder potential, we obtain the average conductance $\langle G\rangle$ and the variance ${\rm Var}\,G=\langle G^{2}\rangle-\langle G\rangle^{2}$. Results are shown in Fig.\ \ref{fig_mu00} at the Dirac point ($\mu_{0}=0$) and in Fig.\ \ref{fig_mu05} at $\mu_{0}=\tau/2$.

The Altshuler-Lee-Stone theory of universal conductance fluctuations (UCF) gives a variance \cite{Alt85,Lee85,Akk06}
\begin{equation}
{\rm Var}\,G_{\rm UCF}=C\,\frac{1}{\beta}\left(\frac{se^{2}}{h}\right)^{2}\frac{W}{L},\;\;{\rm if}\;\;W\gg L,\label{VarG}
\end{equation}
with $C=(3/\pi^{3})\zeta(3)=0.116$ and $\zeta(x)$ the Riemann zeta function. For atomic-scale disorder, the symmetry index $\beta=1$ (orthogonal symmetry) and the degeneracy factor $s=2$ (only spin degeneracy). For smooth disorder, one has $\beta=4$ (symplectic symmetry) and $s=4$ (both spin and valley degeneracy). In each case, the variance thus has the same value ${\rm Var}\,G_{\rm UCF}=C\,(W/L)(2e^{2}/h)^{2}$.

In Figs.\ \ref{fig_mu00}b,\ref{fig_mu05}b we see that the conductance fluctuations approach the UCF value for sufficiently strong disorder. This is by itself remarkable, since the Altshuler-Lee-Stone theory requires {\em weak disorder}, such that the conductivity $\sigma\equiv\langle G\rangle L/W\gg e^{2}/h$. Our numerical data for ${\rm Var}\,G$ only approaches ${\rm Var}\,\,G_{\rm UCF}$ when the disorder is so strong that $\sigma\simeq e^{2}/h$. For weaker disorder, the conductance fluctuations first rise to a peak value ${\rm Var}\,G_{\rm peak}$ well above ${\rm Var}\,G_{\rm UCF}$, and then drop to zero upon entering the ballistic regime. 

The increase of the conductance fluctuations above the UCF value does not happen for the Anderson model of atomic-scale disorder (solid and dashed curves). For smooth disorder the enhancement factor ${\rm Var}\,G_{\rm peak}/{\rm Var}\,G_{\rm UCF}$ increases with increasing Fermi energy $\mu_{0}$ --- it is therefore not restricted to the vicinity of the Dirac point. The enhancement factor also increases with increasing $\xi$ (not shown), but at fixed $\xi$ it is insensitive to the system size (compare open and filled symbols in Figs.\ \ref{fig_mu00}b,\ref{fig_mu05}b). The anomalous enhancement does not, therefore, appear to be a finite-size effect. 

The transport mean free path (\ref{ltrdef}) at $\mu_{0}=\tau/2$ is $l_{\rm tr}=4\sqrt{3}\,a/K_{0}$ (for smooth disorder), so $l_{\rm tr}/L\approx 0.1$ at the peak of maximal conductance fluctuations in the largest system considered. We are therefore well outside of the ballistic regime, but the UCF value characteristic of diffusion is not reached until the mean free path has been reduced by another factor of ten. By comparing the data in Fig.\ \ref{fig_mu05}a for $L/a=40$ and $L/a=72$, we can conclude that the diffusive regime (with a scale invariant conductivity) is not reached until $K_{0}\gtrsim 10$ for smooth disorder, while the diffusive regime is reached already for $K_{0}\gtrsim 0.1$ for atomic-scale disorder.\footnote{We have calculated the third and fourth cumulants, to search for deviations from a Gaussian conductance distribution. In the parameter range of Fig.\ \ref{fig_mu05} no significant deviations are obtained in the case of smooth disorder. We need atomic-scale disorder to obtain significantly non-Gaussian distributions at large disorder strengths.}

While the disappearance of the anomaly for atomic-scale disorder unambiguously indicates that the symplectic symmetry of the Dirac Hamiltonian is responsible for it, we have not been able to explain our simulations consistently in terms of existing transport theories for Dirac fermions \cite{Per06,Che06,Zie06,Nom06,Ale06,Ost06,Hwa06,Ryu06,Tit06}. Certain features of the data suggest a partial explanation.

First of all, at the Dirac point ($\mu_{0}=0$), the enhancement of the conductance fluctuations happens in the same range of disorder strengths as the enhancement of the conductivity above the ballistic value \cite{Kat06,Two06}
\begin{equation}
\sigma_{\rm ballistic}=\frac{4}{\pi}\frac{e^{2}}{h}.\label{sigmaball}
\end{equation}
This increase of $\sigma$ was predicted by Titov \cite{Tit06} as a manifestation of resonant transmission of evanescent modes. We would expect such transmission resonances to enhance the mesoscopic fluctuations, but we would also expect the effect to diminish as the evanescent modes become propagating away from the Dirac point. Instead, the peak in ${\rm Var}\,G$ becomes larger with increasing $\mu_{0}$, while the peak in $\sigma$ disappears.

A second striking feature of the numerical data is that an increase of the impurity density $N_{\rm imp}/N_{\rm tot}$ and a decrease of the impurity potential $\delta$ at fixed $K_{0}$ has no significant effect on the conductance (compare the different open symbols in Figs.\ \ref{fig_mu00},\ref{fig_mu05}, which all lie approximately on a single curve). This signifies that the transition from the anomalously large fluctuations at weak disorder to the UCF value at stronger disorder is not related to the Born-Unitarity transition of Ref.\ \cite{Ost06} (which should appear at smaller $K_{0}$ for smaller $N_{\rm imp}/N_{\rm tot}$).

The percolation transition of Ref.\ \cite{Hwa06} is more likely to be at the origin of the strong increase of the conductance fluctuations away from the Dirac point (where $k_{F}\xi\gtrsim 1$), in the regime $0.1\lesssim K_{0}\lesssim 10$ in between the ballistic and diffusive transport regimes. One would expect the presence or absence of a percolating trajectory to produce large sample-to-sample fluctuations in the conductance, which would increase both with increasing $k_{F}$ and with increasing $\xi$ --- as observed in our simulations. This interpretation would imply that the conductance fluctuations result from variations in {\em trajectories\/} rather than fluctuations in {\em phase shifts}.

\begin{figure}[tb]
\centerline{\includegraphics[width=0.9\linewidth]{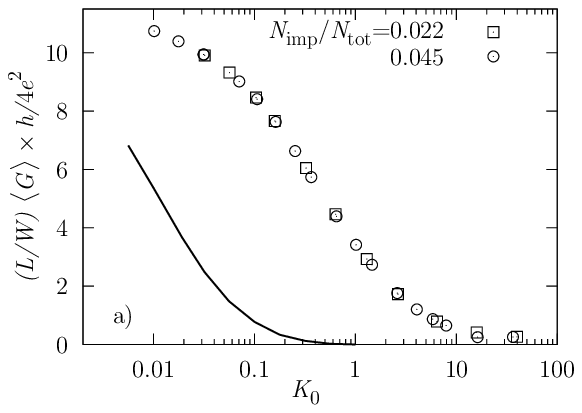}}

\centerline{\includegraphics[width=0.9\linewidth]{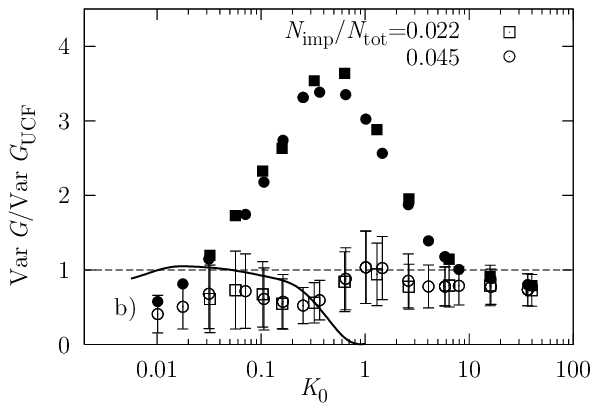}}

\caption{\label{fig_WL1}
Average and variance of the conductance away from the Dirac point ($\mu_{0}=\tau/2$) as a function of disorder strength, for a square sample ($L=W=121\,a$). The filled symbols in (b) give the variance ${\rm Var}\,G$ of the sample-to-sample fluctuations, while the open symbols give the variance ${\rm Var}_{\mu}\,G$ of the energy-dependent fluctuations. (The latter quantity was calculated for a given sample by varying $\mu_{0}\in(0.44\,\tau,0.5\,\tau)$ and then averaging the resulting variance over 400 samples.) The solid curves in (a) and (b) represent, respectively, $\langle G\rangle$ and ${\rm Var}\,G$ in the Anderson model of atomic disorder. (In that model there is no significant difference between ${\rm Var}\,G$ and ${\rm Var}_{\mu}\,G$.)
}
\end{figure}

To support this interpretation we compare in Fig.\ \ref{fig_WL1} the variance ${\rm Var}\,G$ of the sample-to-sample fluctuations with the variance ${\rm Var}_{\mu}\,G$ of the fluctuations obtained in a given sample upon varying the Fermi energy $\mu_{0}$ over a narrow interval. The former quantity contains contributions both from variations in trajectories and variations in phase shifts, while in the latter quantity variations in phase shifts give the dominant contribution. To improve the numerical efficiency we took $W/L=1$ for this comparison. The results for ${\rm Var}\,G$ (filled symbols in Fig.\ \ref{fig_WL1}b) are similar to those plotted in Fig.\ \ref{fig_mu05}b for $W/L=3$: A large enhancement appears of the sample-to-sample fluctuations above the UCF value. In contrast, the variance ${\rm Var}_{\mu}\,G$ of the energy-dependent fluctuations (open symbols) does {\em not\/} show this enhancement, instead agreeing well with the UCF prediction [which for $W/L=1$ equals ${\rm Var}\,G_{\rm UCF}=0.186\times(2e^{2}/h)^{2}$].

In the Altshuler-Lee-Stone theory of UCF one has ${\rm Var}\,G={\rm Var}_{\mu}\,G$: Sample-to-sample fluctuations and fluctuations as a function of energy or magnetic field give the same variance. Our computer simulations imply that, remarkably enough, this socalled {\em ergodicity\/} of the mesoscopic fluctuations does not hold in graphene. An analytical theory to explain this unexpected numerical result is still lacking.

\acknowledgments
We thank J. H. Bardarson, V. I. Falko, G. Montambaux, and M. Titov for valuable discussions and correspondence. This research was supported by the Dutch Science Foundation NWO/FOM and by the European Community's Marie Curie Research Training Network (contract MRTN-CT-2003-504574, Fundamentals of Nanoelectronics). AR acknowledges support by the Polish Ministry of Science (Grant No.\ 1-P03B-001-29) and by the Polish Science Foundation (FNP).

\end{document}